\documentclass [12pt,nofootinbib,eqsecnum,amsfonts,amssymb,aps]{revtex4}

\input epsf

\usepackage[usenames]{color}
\usepackage{graphicx}
\usepackage{bm}
\usepackage{rotating}

\newcommand{\beq}{\begin{equation}}
\newcommand{\eeq}{\end{equation}}
\newcommand{\beqs}{\begin{eqnarray}}
\newcommand{\eeqs}{\end{eqnarray}}

\begin{document}

\baselineskip 6.0mm

\title{On General-$n$ Coefficients in Series Expansions for
  Row Spin-Spin Correlation Functions in the Two-Dimensional Ising Model}

\author{Robert Shrock}

\affiliation{C. N. Yang Institute for Theoretical Physics and
Department of Physics and Astronomy, \\
Stony Brook University, Stony Brook, New York 11794, USA \footnote{
email: robert.shrock@stonybrook.edu, orcid: 0000-0001-6541-0893} \\ 
keywords: Ising correlation functions, series expansions}

\begin{abstract}

  We consider spin-spin correlation functions for spins along a row,
  $R_n = \langle \sigma_{0,0}\sigma_{n,0}\rangle$, in the
  two-dimensional Ising model. We discuss a method for calculating
  general-$n$ expressions for coefficients in high-temperature and
  low-temperature series expansions of $R_n$ and apply it to obtain
  such expressions for several higher-order coefficients.  In addition
  to their intrinsic interest, these results could be useful in the
  continuing quest for a nonlinear ordinary differential equation whose
  solution would determine $R_n$, analogous to the known nonlinear ordinary
  differential equation whose solution determines the diagonal
  correlation function $\langle \sigma_{0,0}\sigma_{n,n}\rangle$ in
  this model.

\end{abstract}

\maketitle

\newpage

\pagestyle{plain}
\pagenumbering{arabic}


\section{Introduction}
\label{intro_section}

Spin-spin correlation functions contain information about the degree
of magnetic ordering in a spin model. The two-dimensional Ising model
provides a valuable context in which one can obtain exact closed-form
analytic expressions for these correlation functions.  In thermal
equilibrium at temperature $T$ on the square lattice, the partition
function of the (zero-field, isotropic, spin-1/2, nearest-neighbor)
Ising model is given by
\beq
Z = \sum_{\{ \sigma \} } e^{-\beta {\cal H}} \ , 
\label{z}
\eeq
where the Hamiltonian is  
\beq
{\cal H} = -J \sum_{nn} \sigma_{\vec r} \sigma_{{\vec r}'} \ .
\label{ham}
\eeq
In Eq. (\ref{ham}), $\beta = 1/(k_BT)$; the sum is over
nearest-neighbor ($nn$) sites on the lattice; and $\sigma_{\vec r} =
\pm 1$ is a classical spin variable defined on each lattice
site. Given the well-known mapping on a bipartite lattice between the
ferromagnetic ($J > 0$) and antiferromagnetic ($J < 0$) spin-spin
couplings \cite{domb}, one can, with no loss of generality, take $J >
0$, and we will do this. In the following, we assume the thermodynamic
limit.  This model has a global ${\mathbb Z}_2$ symmetry, which is
spontaneously broken with the onset of a nonzero spontaneous
magnetization $M$ as the temperature decreases below the critical
temperature, $T_c$. The system undergoes a continuous, second-order
phase transition at this critical temperature.

The two-dimensional Ising model has the appeal that many of its
properties are known exactly.  The free energy was calculated by
Onsager \cite{o}, and an expression for the magnetization was first
published by Yang \cite{ym}. A method for calculating spin-spin
correlation functions in terms of Toeplitz determinants was developed
by Kaufman and Onsager \cite{ko} and later extended by Montroll,
Potts, and Ward \cite{mpw}.  Some reviews of the Ising model include
\cite{domb,mwbook}.  The critical behavior is known exactly; the
thermal and magnetic exponents are $y_t=1$ and $y_h=15/8$, and hence
the critical exponents $\nu$, $\alpha$, $\beta$, $\gamma$, etc., are
known for models in the $d=2$, ${\mathbb Z}_2$ universality class of
second-order phase transitions, of which the nearest-neighbor spin-1/2
Ising model is arguably the simplest example. Generalizations to
anisotropic couplings, spin $s \ge 1$, and non-nearest-neighbor two-spin
and multi-spin interactions preserving the ${\mathbb Z}_2$ symmetry
that do not cause frustration are in the same universality class. The
critical behavior was elucidated in the context of continuous spatial
dimensionality (above the lower critical dimensionality, $d=1$) via
the momentum-space renormalization group and $\epsilon$ expansion, where
$\epsilon=4-d$ \cite{wilson_kogut,fisher_rmp,zinnjustin80}.  Further
insight into the critical behavior of the $d=2$, ${\mathbb Z}_2$
universality class was obtained by means of conformal algebra methods
\cite{bpz}, which showed that the critical behavior is described by a
rational conformal field theory with central charge $c=1/2$ (reviewed,
e.g., in \cite{cftbook}). Although the 2D Ising model is classical, it
can be related to a 1D quantum spin chain \cite{suzuki}.  In addition
to these works, some other studies relevant to spin-spin correlation
functions in the two-dimensional Ising model include
\cite{stephenson}-\cite{mm_nu}.

While much is thus known about the two-dimensional Ising model,
there are still interesting aspects to study. Among these are various
properties of the spin-spin correlation functions.  We denote the
spin-spin correlation function as $\langle \sigma_{\vec r}
\sigma_{{\vec r}'} \rangle$, where ${\vec r}$ and ${\vec r}'$ are
sites on the square lattice and $\langle {\cal O}\rangle$ denotes the thermal
average of an operator ${\cal O}$. Given the homogeneity of the square
lattice, one can, with no loss of generality, take one spin to be
located at the origin and thus consider
\beq
\langle \sigma_{\vec 0} \sigma_{\vec r}\rangle \equiv C({\vec r}) \ .
\label{c}
\eeq
We write $\vec r = (m,n)$ so that $C({\vec r}) \equiv C(m,n)$. 
The spin-spin correlation function for two spins along a row is 
\beq
R_n \equiv C(n,0) \equiv \langle \sigma_{0,0} \sigma_{n,0}\rangle \ .
\label{rn}
\eeq
From the isotropy of the spin-spin couplings in (\ref{ham}), it
follows that the correlation functions for equidistantly separated
spins along a row and column are equal: $C(n,0)=C(0,n)$.  We denote
the correlation function for spins along a diagonal of the lattice as
\beq
D_n \equiv C(n,n) \equiv \langle \sigma_{0,0} \sigma_{n,n}\rangle \ .
\label{dn}
\eeq
Note the symmetry relation $C(n,n)=C(n,-n)$. In studying spin-spin
correlation functions of the two-dimensional Ising model, one
acknowledges that these are not universal in the sense of the
renormalization group; that is, modifications of the model
(\ref{z})-(\ref{ham}) such as the generalization to spin $s \ge 1$ and/or
addition of (nonfrustrating) non-nearest-neighbor spin-spin or
multispin interactions preserving the ${\mathbb Z}_2$ symmetry would
change $\langle \sigma_{0,0}\sigma_{m,n}\rangle$ without changing the
universality class of the phase transition.  Nevertheless, these
correlation functions contain useful information about the behavior of
the model. We define the following notation: 
\beq
K = \beta J \ , \quad v = \tanh K \ , \quad x = v^2 \ , \quad 
z = e^{-2K} \ , \quad u = z^2 = e^{-4K} \ .
\label{variables}
\eeq
Correlation functions are commonly expressed as functions of the variables
\beq
k_> = \sinh^2(2K) \ , \quad\quad
k_< = \frac{1}{k_>} = \frac{1}{\sinh^2(2K)} \ .
\label{kgkl}
\eeq
Recall that (as follows from duality) the critical point occurs at
$v_c=z_c=\sqrt{2}-1$, i.e., $K_c = J/(k_BT_c) = (1/2)\ln(\sqrt{2}+1)$,
at which point $k_> = k_< = 1$.  The high-temperature (HT) series
expansions of spin-spin correlation functions are commonly expressed
as series in powers of $v$, while the low-temperature (LT) expansions
on a bipartite lattice such as the square lattice considered here, are
series in powers of $u$.

In Ref. \cite{jimbo_miwa}, Jimbo and Miwa showed that $D_n$ can be calculated
in terms of solutions to a (nonlinear, second-order) ordinary differential
equation (ODE) of Painlev\'e VI type (see Appendix \ref{dn_htseries_appendix}).
Subsequently, there has been a quest to find an analogous nonlinear ordinary
differential equation whose solutions would yield the general spin-spin
correlation function $C(m,n)$ in this model. However, as emphasized recently in
\cite{mm_nu}, this is still an open problem.  Even for $R_n$, to our
knowledge, such a generalization of the Jimbo-Miwa ODE has not been
found. Indeed, in the absence of an existence proof, it is not clear if
such a (nonlinear, second-order) ODE whose solutions would yield the $R_n$, 
analogous to the Jimbo-Miwa Painlev\'e VI ODE for $D_n$ (see Appendix A), 
exists. Investigations into this can make use of exact calculations of
correlation functions. The $D_n$ and $R_n$ can be expressed as Toeplitz
determinants, and this method was used in \cite{diag,row} to calculate these
correlation functions for $n$ up to 6 and to present exact expressions for $n$
up to 5. Exact calculations of some other $C(m,n)$ were given in
\cite{offaxis}.  It was shown in \cite{diag} that $D_n$ is a homogeneous
polynomial of degree $n$ in the complete elliptic integrals $K(k)$ and $E(k)$,
where $k=k_>$ for $T \ge T_c$ and $k=k_<$ for $T \le T_c$.  The general
structure of $R_n$ for the model of Eqs. (\ref{z}), (\ref{ham}) was determined
in \cite{row} and is substantially more complicated, as will be reviewed
below. As shown in \cite{ayp02}, the $C(m,n)$ for this model can be efficiently
calculated recursively using certain quadratic relations \cite{perk80} together
with some initial inputs.  Both of these methods yield specific correlation
functions, e.g., $R_6$, $R_7$, etc. for higher $n$.  In searching for a
nonlnear ODE for $R_n$ analogous to the Jimbo-Miwa ODE for $D_n$, it would be
convenient to use inputs that are general functions of $n$, rather than having
to recursively compute $R_n $ for successive fixed values of $n$.  For this
purpose, high-temperature and low-temperature series expansions can be useful,
if one knows general-$n$ expressions for the coefficients.  However, standard
procedures for calculating these series expansions are based on enumeration of
graphs for a given correlation function $C(m,n)$ and, except for the first or
second leading terms, do not normally yield expressions that are general
functions of $(m,n)$.  Here we focus on $R_n$.  The leading term in the
high-temperature series expansion of $R_n$ is $v^n$, and an elementary
graphical enumeration yields the first higher-order term as $n(n+1)v^{n+2}$,
but we are not aware of general-$n$ expressions for still higher-order terms in
the literature. Similar comments apply for the low-temperature series expansion
of this correlation function.

In this paper we shall discuss an approach that can yield general-$n$
coefficients of higher-order terms in high-temperature and low-temperature
expansions of the row correlation functions $R_n$ for the two-dimensional Ising
model defined by Eqs. (\ref{z})-(\ref{ham}) on the square lattice.  Our
procedure makes use of exact calculations of individual $R_n$. We illustrate
the approach by computing general-$n$ coefficients of several higher-order
terms in high-temperature expansions of $R_n$ and low-temperature expansions of
$(R_n)_{\rm conn.}$. In addition to their intrinsic interest, this method and
these results should be useful in the continuing endeavor to find a nonlinear
ordinary differential equation for $R_n$ analogous to the one derived for $D_n$
by Jimbo and Miwa in \cite{jimbo_miwa}.  Our work here is complementary to
studies of form factor expansions for Ising correlation functions (e.g.,
\cite{wmtb,painleve_fuchs,fuchs_painleve,lm,guttmann_formfactors, chi6}). It is
also complementary to studies of properties of the Ising model susceptibility
$\chi$ (e.g., \cite{sgmme73,guttmann_series,chisq,chi01,chin,chi6}), since the
latter involves a sum over all connected spin-spin correlation functions, not
just $R_n$, via the relation $\beta^{-1} \chi = \sum_{\vec r} C(\vec r)_{\rm
  conn.}$.

This paper is organized as follows.  In Section
\ref{rcf_structure_section} we review the general structural form for
$R_n$ obtained in \cite{row}.  In Sections
\ref{high_temp_series_section} and \ref{low_temp_series_section} we
use the exactly calculated $R_n$ from \cite{row} to infer general-$n$
expressions for several coefficients of higher-order terms in
high-temperature series for $R_n$ and low-temperatures series for
$(R_n)_{\rm conn.}$. Our conclusions are given in Section
\ref{conclusion_section}.  Some related results are included in appendices.  


\section{Structure of Row Correlation Functions}
\label{rcf_structure_section}

From our analysis in \cite{row}, we inferred the following general
structural form for the row correlation functions $R_n$. These have
different analytic forms $R_{n,+}$ and $R_{n,-}$ for $T > T_c$ and $T
< T_c$, respectively (which are equal at $T_c$):

\beq
 n \ {\rm even}: \quad 
R_{n,\pm} = B_n \, k^{-p_n} \sum_{\ell=0}^{n/2} \pi^{-2\ell} \,
\sum_{s=0}^{2\ell} {\cal R}^{(n,\pm)}_{2\ell-s,s}(k) \,E(k)^{2\ell-s}
\bar K(k)^s \ , 
\label{rcf_even_n}
\eeq
where $k=k_>$ for $T \ge T_c$ and $k=k_<$ for $T \le T_c$ and
\beqs
&& n \ {\rm odd}, \ T \ge T_c:  \cr\cr
&& R_{n,+} = B_n \, k_>^{-p_n} (1+k_>^{-1})^{1/2}
\sum_{\ell=0}^n \pi^{-\ell} \sum_{s=0}^\ell
{\cal R}^{(n)}_{\ell-s,s}(k_>) \, (k_>-1)^{(1/2)[1-(-1)^\ell]\delta_{s,0}}
\, E(k_>)^{\ell-s} \bar K(k_>)^s \cr\cr
&&
\label{rcf_odd_n_high}
\eeqs
\beqs
&& n \ {\rm odd}, \ T \le T_c:  \cr\cr
&& R_{n,-} = B_n \, k_<^{-p_n} (1+k_<)^{1/2}
\sum_{\ell=0}^n (-\pi)^{-\ell} \sum_{s=0}^\ell
{\cal R}^{(n)}_{\ell-s,s}(k_<) \, \, (k_<-1)^{(1/2)[1-(-1)^\ell]\delta_{s,0}}
\, E(k_<)^{\ell-s} \bar K(k_<)^s \ . \cr\cr
&&
\label{rcf_odd_n_low}
\eeqs
In Eqs. (\ref{rcf_even_n})-(\ref{rcf_odd_n_low}), $B_n$ is a numerical
prefactor; $p_n$ is an integer power\footnote{
The quantities $B_n$ and $p_n$ and the dummy index $s$ were denoted $D_n$,
$q_n$, and $r$ in \cite{row}; here we relabel these to avoid confusion with 
our notation $D_n$ for $C(n,n)$ and $r = |\vec r|$.}
and we define the compact notation 
\beq
\bar K(k) \equiv (k-1)K(k) \ . 
\label{kbar}
\eeq
The first
five $B_n$ were given in \cite{row}, viz., $B_1=B_2=B_3=1$,
$B_4=1/(3^2)$, and $B_5 = 1/(3^4)$ \cite{row}. The sixth is
$B_6=1/(3^6 \cdot 5^2)$.  The values of the power $p_n$ in
Eqs. (\ref{rcf_even_n})-(\ref{rcf_odd_n_low}) were listed (denoted as
$q_n$) for $n$ up to 5 in \cite{row}.  Here we observe that for the
known $p_n$ with $1 \le n \le 6$, the values are consistent with the
general formula
\beq
p_n = \Big [ \frac{n^2}{4} \Big ]_{\rm floor} \ , 
\label{pn}
\eeq
where for $\nu \in {\mathbb R}$, $[\nu]_{\rm floor}$ is the greatest integer
$\le \nu$. These powers $p_n$ are the same as the powers that occur in the 
general structural form for the diagonal correlation function $D_n$ that
we found in \cite{diag}. 
In the remainder of the paper we will sometimes suppress the subscripts
$\pm$ in the notation, with it being understood implicitly that
$R_n \equiv R_{n,+}$ for $T \ge T_c$ and
$R_n \equiv R_{n,-}$ for $T \le T_c$.
Note that although $K(k)$ is logarithmically divergent as $k
\nearrow 1$, this divergence is removed by the prefactor $(k-1)$ in $\bar
K(k)$. Indeed,
\beq
\lim_{k \to 1}\bar K(k)=0 \ , 
\label{limkk1}
\eeq
although the derivative $(d/dk) \bar K(k)$ is logarithmically
divergent as $k \to 1$.

For a given $n$, the terms in $R_n$ can be divided into sets such that
all of the terms in each set are homogeneous polynomials in $E(k)$ and
$\bar K(k)$ of a given degree.  We label this degree as the ``level''
of the set.  For $R_{n,\pm}$ with odd $n$, as is evident from
Eqs. (\ref{rcf_odd_n_high}) and (\ref{rcf_odd_n_low}), these terms are
explicitly of the form $E(k)^{\ell-s} \bar K(k)^s$ with $\ell$ in the
range $0 \le \ell \le n$ and, for a given $\ell$, with $s$ in the
range $0 \le s \le \ell$, where $k=k_>$ for $T \ge T_c$ and $k=k_<$
for $T \le T_c$. The corresponding coefficients ${\cal
  R}^{(n)}_{\ell-s,s}$ in $R_{n,\pm}$ are polynomials in the
respective $k$ elliptic modulus variables. For even $n$, only
even-degree levels occur, running over $2\ell = 0, 2,...,n$, as is
evident in Eq. (\ref{rcf_even_n}).  Another difference between the
$R_n$ with even and odd $n$ is that for odd $n$, the same coefficient
polynomial ${\cal R}^{(n)}_{\ell-s,s}(k)$ occurs for $T \ge T_c$ and
$T \le T_c$ with the respective assignments $k=k_>$ and $k=k_<$,
whereas for even $n$, the ${\cal R}^{(n,+)}_{2\ell-s,s}(k_>)$ and
${\cal R}^{(n,-)}_{2\ell-s,s}(k_<)$ are different functions of their
respective arguments, $k_>$ and $k_<$. A third difference between the
$R_n$ for even and odd $n$ is that the $R_n$ for odd $n$ contain a
square-root prefactor, $(1+k_>^{-1})^{1/2} = (1+k_<)^{1/2}$, whereas
the $R_n$ for even $n$ do not contain such square-root prefactors
\footnote{We note some misprints in \cite{mpw}, \cite{diag}, and
  \cite{row}.  In Eq. (A19) of \cite{mpw}, the expression
  $-\frac{1}{2}\gamma_2 [F_{m_1,m_2} + F_{m_1,m_2-1}]$ should read
  $-\frac{1}{2}\gamma_2 [F_{m_1,m_2+1} + F_{m_1,m_2-1}]$.  In
  \cite{diag} there was a misprint in the overall sign of ${\cal
    P}^{(5,-)}_{2,3}$, which should be reversed.  In \cite{row}, the
  coefficient ${\cal R}^{(4,-)}_{3,1}$ should be multipled by
  $(k_<-1)$.}.

The general structural form that we inferred for $R_n$ is considerably
more complicated than the form that we had found in \cite{diag} for
$D_n$. These have different expressions $D_{n,+}$ and $D_{n,-}$
for $T > T_c$ and $T < T_c$ (which are equal at $T=T_c$):
\beq
D_{n,\pm} = A_n \pi^{-n} \, k^{-2p_n-[1-(-1)^n]\Theta(T-T_c)/2} \,
\sum_{s=0}^{n} {\cal P}^{(n,\pm)}_{n-s,s}(k) \, 
(k^2-1)^{\delta_{s,1}\Theta(T_c-T)} \, E(k)^{n-s} \, [(k^2-1)K(k)]^s \ , 
\label{dcf}
\eeq
where again $k=k_>$ if $T \ge T_c$ and $k=k_<$ if $T \le T_c$; and $\Theta(x)$
is the Heaviside step function, defined as $\Theta(x)=1$ if $x > 0$ and
$\Theta(x)=0$ if $x \le 0$.  One of the most striking differences is that $D_n$
is a homogeneous polynomial of degree $n$ in $E(k)$ and $K(k)$, while $R_n$ has
the multi-``level'' structure of Eqs.  (\ref{rcf_even_n}),
(\ref{rcf_odd_n_high}), and (\ref{rcf_odd_n_low}).  Furthermore, calculations
for $D_n$ and the structural form presented in \cite{diag} apply for the
general anisotropic case $J_1 \ne J_2$, with $k_> = k_<^{-1} =
\sinh(2K_1)\sinh(2K_2)$ and $K_i=\beta J_i$, whereas in the anisotropic case,
other correlation functions such as $R_n$ would involve not just complete
elliptic integrals of the first and second kinds, but also those of the third
kind, as was already evident for $R_1$ \cite{mwbook,mm_anisotropic}. Our
methods could be applied to this case in future work, although the series
expansions would depend on two variables, e.g., $v_i = \tanh K_i$ where $K_i =
\beta J_i$, $i=1,2$, or equivalently, $s_i = \sinh(2K_i)$ with $i=1,2$ for the
high-temperature expansions, and similarly on $1/s_i$ with $i=1,2$ for the
low-temperature expansions.


\section{General-$n$ Coefficients in the High-Temperature Series Expansion
  of $R_n$}
\label{high_temp_series_section}

Here we report our new results on general-$n$ coefficients of higher-order
terms in the high-temperature Taylor series expansion of $R_n$. This is
analogous to the calculation of general-$n$ coefficients in the HT expansion of
$D_n$ in \cite{ode} (see Appendix \ref{dn_htseries_appendix}), with the crucial
difference that for $D_n$ we were able to make use of the fact that the $D_n$
can be determined in terms of solutions of the Painlev\'e VI ODE
\cite{jimbo_miwa}, whereas here no analogous (nonlinear) ODE for $R_n$ is
known. Hence, we make use of the $R_n$ calculated in \cite{row}.

The standard procedure for calculating the high-temperature Taylor
series expansion for $R_n=\langle \sigma_{0,0}\sigma_{n,0}\rangle$
enumerates the contributions from paths on the bonds of the lattice of
minimal length and progressively greater lengths joining the points
$(0,0)$ and $(n,0)$ (e.g., \cite{domb}). The number of bonds in the
path is then the power of $v$ in a given term in the series expansion,
and the coefficient of each term is a positive integer. If one factors
out an overall factor of $v^n$ in the small-$v$ expansion of $R_n$,
the rest of the series is a series in powers of $v^2$.  This is an
elementary consequence of the fact that if one reverses the sign of
the spin-spin coupling $J$, and hence the sign of $K$ and of $v=\tanh
K$, then $R_n \to (-1)^n R_n$.  The lowest-order term, $v^n$, in the
small-$v$ series expansion of $R_n$, arises from the unique graph
consisting of a straight path from $(0,0)$ to $(n,0)$, of length $n$.
Thus, the high-temperature expansion for $R_n$ has the general form
\beq
R_n = v^n \Big [ 1 + \sum_{j=1}^\infty r_{n,2j} \, v^{2j} \Big ] \ ,
\label{rn_hightemp_series}
\eeq
where the $r_{n,2j}$ are positive integers.  Aside from the leading
$v^n$ term, the paths contributing to all higher-order terms include
bonds above the direct, horizontal path, and corresponding paths that
are related to these by reflection about the horizontal axis.  That
is, for each path including bonds above the direct route along the
horizontal axis joining the sites $(0,0)$ to $(n,0)$, there is a path
that is obtained by this reflection process.  Therefore, the
$r_{n,2j}$ are even integers.

To begin, we discuss the graphical derivation of the first subleading
term in Eq. (\ref{rn_hightemp_series}), namely, $r_{n,2}v^{n+2}$. This
term arises from paths of length $n+2$ bonds connecting the sites
$(0,0)$ and $(n,0)$. An elementary enumeration counts these paths.
The sites $(0,\ell)$ with $0 \le \ell \le n$ comprise $n+1$ vertices
on the square lattice.  One set of paths of length $n+2$ connecting
$(0,0)$ and ($n,0)$ involves a $90^\circ$ turn upwards at one of these
$n+1$ sites followed by a continuation along horizontal bonds, and
then a $90^\circ$ turn downward and final continuation to the point
$(n,0)$.  For each such path, there is also a corresponding path
obtained by reflecting about the horizontal axis, so that the first
right-hand turn is downward instead of upward. There are ${n+1 \choose
  2}$ paths in the first set, and hence $2{n+1 \choose 2} = n(n+1)$
paths of length $n+2$ joining the points $(0,0)$ and $(n,0)$.  This
simple combinatoric argument yields the coefficient, $r_{n,2}$, of the
$v^{n+2}$ term in the high-temperature expansion of $R_n$, namely
\beq
r_{n,2}=n(n+1) \ .
\label{rn_j2}
\eeq
This is manifestly even, since either $n$ or $n+1$ is even.

Now from high-temperature series expansions of our calculations of
$R_n$ for $n$, we determine the following general-$n$ expression for
the next-to-next-leading-order coefficient, $r_{n,4}$.  For reference,
we list these expansions for $n$ up to 6 in Appendix
\ref{rn_htseries_appendix}.  Our method is motivated by the structural
form as polynomials in $n$ that we obtained for coefficients of
higher-order terms in the HT series for $D_n$ in \cite{ode} (reviewed
in Appendix \ref{dn_htseries_appendix}).  We thus fit the respective
$O(v^{n+4})$ terms in the HT series expansions of the exact
expressions for $R_n$ to a polynomial.  This is an overconstrained
fit, and we obtain the result
\beq
r_{n,4} = \frac{n}{4}(n^3+2n^2+3n+10) \ . 
\label{rn_j4}
\eeq
Although there is an extensive literature on series expansions of quantities in
the two-dimensional Ising model, we are not aware of this expression for
$r_{n,4}$ having appeared in this literature. An alternate approach to
determining $r_{n,4}$ would make use of an enumeration of all graphs that
contribute to the $O(v^{n+4})$ term in the high-temperature expansion of $R_n$
for arbitrarily great $n$. This result is given to illustrate the method;
clearly, one could proceed to calculate coefficients of more higher-order
terms, $r_{n,6}$, etc.  Since the $r_{n,2j}$ with $2j \ge 6$ are higher-degree
polynomials in $n$, the procedure for calculating these polynomials via
overconstrained fits requires the input of a larger number of row correlation
functions.  However, as emphasized, the value of this method is that the
resultant coefficient applies for general $n$ and hence is directly applicable
to the search for a nonlinear differential equation whose solution would yield
$R_n$.

Despite the prefactor of 1/4, it is easy to show that the expression
for $r_{n,4}$ in Eq. (\ref{rn_j4}) is an integer, and, furthermore, is
even.  This is proved by induction, starting from any of the known
$r_{n,4}$ values for $1 \le n \le 6$, each of which is even.  Given
that there exists an $n$ such that $r_{n,4}$ is even, to carry out the
inductive proof, one must prove that $r_{n+1,4}$ is also even.  This
can be done by showing that the difference, $r_{n+1,4}-r_{n,4}$, is
even, i.e., $r_{n+1,4}-r_{n,4}=2p$ for some (positive) integer $p$.
We calculate
\beq
r_{n+1,4}-r_{n,4} = (n+2)(n^2+n+2) \ .
\label{rnp1_rn_diff}
\eeq
Since the factor $(n+2)$ can be even or odd, we thus need to show that
$n^2+n+2$ is always even. This follows directly by observing that
$n^2+n+2=n(n+1)+2$.  Now $n(n+1)$ is manifestly even, since either $n$
or $n+1$ is even, and hence $n(n+1)+2$ is even. This completes the proof
that the expression for $r_{n,4}$ in Eq. (\ref{rn_j4}) is an even
(positive) integer.  

One can also express these results equivalently as series expansions in powers
of the variable $\sqrt{k_>}$, using the relation (\ref{kgval}). It is
convenient to introduce the variable $\hat k_> = (1/4)k_>$ as in
Eq. (\ref{khat}). Then Eq. (\ref{rn_hightemp_series}) can be written as
\beq
R_n = \hat k_>^{n/2} \Big [ 1 + \sum_{\ell=1}^\infty \tilde r_{n,\ell} \, 
\hat k_>^\ell \Big ] \ , 
\label{rnkg_hightemp_series}
\eeq
where 
\beq
\tilde r_{n,1} = n^2 
\label{rntilde_ell1}
\eeq
and
\beq
\tilde r_{n,2} = \frac{1}{4}n(n-1)(n^2-n-8) \ . 
\label{rntilde_ell2}
\eeq
While the coefficients $r_{n,2}$, $r_{n,4}$, and $\tilde r_{n,1}$ are positive
and monotonically increasing as functions of $n$ in the interval $n \ge 1$, the
behavior of $\tilde r_{n,2}$ is more complicated.  As a function of $n$, with
$n$ generalized from integral values to real values in this interval $n \ge 1$,
$\tilde r_{n,2}$ decreases from zero at $n=1$ through negative values, reaching
a minimum of $-4$ at $n=(1/2)(1+\sqrt{17}) = 2.56155$ and then increases
monotonically for larger $n$, passing through zero again at
$n=(1/2)(1+\sqrt{33}) = 3.37228$. Thus, $\tilde r_{n,2}$ is negative for $n=2$
and $n=3$, taking the values $\tilde r_{2,2}=\tilde r_{3,2}=-3$. 


\section{General-$n$ Coefficients of Higher-Order Terms in the
  Low-Temperature Series Expansion of $(R_n)_{\rm conn.}$}
\label{low_temp_series_section}

In addition to its intrinsic interest, the spin-spin correlation function
$C(\vec r)$ is important because its limit as $r \to \infty$ determines
the (square of the) spontaneous magnetization: 
\beq
\lim_{r \to \infty} C({\vec r}) = M^2 \ , 
\label{msqrel}
\eeq
where $r \equiv |\vec r|$.  The connected correlation function is then
\beq
C({\vec r})_{\rm conn.} \equiv C({\vec r}) - M^2 \ .
\label{conn}
\eeq
For $T<T_c$ where the spontaneous magnetization is nonzero, an
interesting question concerns the approach to the limit (\ref{msqrel}),
For a given ${\vec r}$, a quantitative measure of this
approach is provided by the ratio 
\beq
A_{C({\vec r})} = \frac{C({\vec r})}{M^2} =
1 + \frac{C({\vec r})_{\rm conn.}}{M^2} \ , 
\label{aratio}
\eeq

In this section, we present our results on general-$n$ coefficients of
higher-order terms in the low-temperature Taylor series expansions of
$(R_n)_{\rm conn.}=R_n-M^2$ and $A_{R_n}=R_n/M^2$. In calculating
$(R_n)_{\rm conn}$, we make use of the
result first published by Yang \cite{ym} for the 
spontaneous magnetization in the two-dimensional Ising model on the
square lattice, 
\beq
M = (1-k_<^2)^{1/8} = \frac{(1+u)^{1/4}(1-6u+u^2)^{1/8}}{(1-u)^{1/2}} \ .
\label{mag}
\eeq
The quantity $M^2$ has the resultant low-temperature Taylor series
expansion 
\beqs
M^2 &=& 1 -4u^2 -16u^3 -64u^4 -272u^5 -1228u^6 -5792u^7 -28192u^8 -140448u^9
\cr\cr
&-& 712276u^{10} - 3663664u^{11} - O(u^{12}) \ .
\label{Msq_taylor}
\eeqs

The property (\ref{msqrel}), together with the property that $M$ and
$R_n$ are continuous functions of $u$, implies that as $n$ increases,
the low-temperature (i.e., small-$u$) Taylor series expansion of $R_n$
must coincide with the small-$u$ expansion of $M^2$ to an increasingly
high order, and the order of the first term in the small-$u$ expansion
of $R_n-M^2$ must go to infinity as $n \to \infty$.  From general
arguments, for an Ising ferromagnet on (the thermodynamic limit of) a
given lattice, $R_n$ is a monotonically decreasing function of $n$
for fixed temperature $T$, i.e., $R_n \ge R_{n+1}$, and hence $R_n \ge
M^2$.  (The two points at which this inequality is realized as an
equality are (i) $T=0$ for any $n$, where $R_n=M^2=1$, and (ii)
$T=\infty$, where for $n \ge 1$, $R_n=M^2=0$.) 

The low-temperature Taylor series expansions of the $R_n$ in powers of $u$
or $k_<$ match the corresponding expansions of $M^2$ to $O(u^{n+1}) = 
O(k_<^{n+1})$ inclusive. Thus, the LT series expansion of $R_n$ has the general form
\beq
(R_n)_{\rm conn.}=4u^{n+2}\Big [1+\sum_{j=1}^\infty \rho_{n,j}u^j \, \Big ] 
          = 4\hat k_<^{n+2}\Big [1+\sum_{j=1}^\infty 
\tilde \rho_{n,j}\hat k_<^j \, \Big ] \ . 
\label{rnconn_smallu}
\eeq
Here it is convenient to use the rescaled variable $\hat k_< = (1/4)k_<$ (as
defined in Eq. (\ref{khat})), since this yields integral coefficients $\tilde
\rho_{n,j}$.  Using LT expansions of the $R_n$ that we have calculated exactly,
we apply the same polynomial fitting procedure that we used for the HT
expansions.  For reference, we list these LT expansions in Appendix
\ref{rn_ltseries_appendix}. We obtain the following general-$n$ expressions for
the $u^{n+3}$ and $u^{n+4}$ terms in Eq. (\ref{rnconn_smallu}):
\beq
\rho_{n,1} = n^2+2n+4 \ , 
\label{rho_n_j1}
\eeq
and
\beq
\rho_{n,2} = \frac{1}{2}(n^4+4n^3+13n^2+26n+32) \ . 
\label{rho_n_j2}
\eeq
Equivalently, for the expansion of $(R_n)_{\rm conn.}$ in terms of 
$\hat k_<$ in Eq. (\ref{rnconn_smallu}), we have
\beq
\tilde \rho_{n,1} = n^2 
\label{rho_tilde_n_j1}
\eeq
and
\beq
\tilde \rho_{n,2} = \frac{1}{2}(n+2)(n^3-2n^2+n+6) \ . 
\label{rho_tilde_n_j2}
\eeq

By combining the LT expansion for $M^2$ with these results, one can thus obtain
the corresponding general-$n$ LT expansion for $R_n$ up to $O(u^{n+4}) = O(\hat
k_<^{n+4})$.  We are not aware of the expressions
(\ref{rho_n_j1})-(\ref{rho_tilde_n_j2}) having appeared before in the
literature.  These results are given to illustrate the method and could be
extended to higher order using additional $R_n$ correlation functions as input.

Note that, despite the prefactor of 1/2, the expression for $\rho_{n,2}$ is an
integer. We give an inductive proof of this. First, this integral property
holds for the LT series for $R_1$. Hence, it is necessary and sufficient to
show that with $\rho_{n,2}$ being integral, so is $\rho_{n+1,2}$.  To do this,
we show that the difference $\rho_{n+1,2}-\rho_{n,2}$ is integral. This
difference is
\beq
\rho_{n+1,2}-\rho_{n,2} = (n+1)(2n^2+5n+11) \ ,
\label{rhon_4_diff}
\eeq
which is obviously integral. The same inductive method shows that 
$\tilde\rho_{n,2}$ is an integer.

Combining these results with the definition $A_{R_n}=R_n/M^2$ yields 
\beqs
A_{R_n} &=& 1 + 4u^{n+2}\Big [ 1 + \rho_{n,1}u + (\rho_{n,2}+4)u^2
  + O(u^3) \Big ] \cr\cr
     &=& 1 + 4u^{n+2}\Big [ 1 + (n^2+2n+4)u +
  \frac{1}{2}(n^4+4n^3+13n^2+26n+40)u^2 + O(u^3) \Big ] \ . \cr\cr
&&
\label{rn_ratio_taylor_general}
\eeqs
Equivalently, in terms of the $\hat k_<$ variable, 
\beqs
A_{R_n} &=& 1 + 4\hat k_<^{n+2}\Big [ 1 + \tilde\rho_{n,1}\hat k_< + 
(\tilde\rho_{n,2}+4)\hat k_<^2 + O(\hat k_<^3) \Big ] \cr\cr
     &=& 1 + 4\hat k_<^{n+2}\Big [ 1 + n^2 \hat k_< +
  \frac{1}{2}(n^4-3n^2+8n+20)\hat k_<^2 + O(\hat k_<^3) \Big ] \ . \cr\cr
&&
\label{rn_ratio_taylor_general2}
\eeqs
%


\section{Conclusions}
\label{conclusion_section}

In this paper we have discussed a method for obtaining general-$n$ expressions
for coefficients of higher-order terms in the high-temperature and
low-temperature series expansions of the spin-spin correlation function $R_n$
in the two-dimensional Ising model on the square lattice and have applied it to
obtain general-$n$ coefficients of several higher-order terms in these series.
This method is complementary to the standard method for calculating these
coefficients, which is via enumeration of graphs that contribute in a given
order of expansion.  It is also complementary to another method that was used
in \cite{ode} for the high-temperature expansions of diagonal correlation
functions $D_n$, which was based on the property that the $D_n$ can be computed
in terms of solutions to the Painlev\'e VI ordinary differential equation of
\cite{jimbo_miwa}.  In addition to the intrinsic interest in the general-$n$
coefficients discussed here, they provide further inputs to the continuing
quest to find a nonlinear ordinary differential equation whose solution would
determine the $R_n$.


\begin{acknowledgments}

  I would like to thank Prof. B. M. McCoy for questions and
  discussions that led me to revisit this subject. I am also grateful
  to Prof. J. W. Essam for an informative communication. This
  research was supported in part by the U.S. National Science
  Foundation Grant NSF-PHY-1915093.

\end{acknowledgments}


\bigskip
\bigskip

\begin{appendix}


\section{General-$n$ Coefficients in the High-Temperature Series
Expansion of $D_n$}
\label{dn_htseries_appendix}

The high-temperature series expansion of $D_n$ in the Ising model
on the square lattice has the form 
\beq
D_n = \sum_{j=n}^\infty c^{(D)}_{n,j} x^j \ ,
\label{dn_high_temp_series}
\eeq
where $x \equiv v^2$ (c.f. Eq. (\ref{variables})) and the subscript
$+$ in $D_{n,+}$ is understood implicitly.  An elementary combinatoric
argument determines the coefficient of the leading-order term as
\beq
c^{(D)}_{n,n} = \frac{(2n)!}{(n!)^2} \ . 
\label{cdnn}
\eeq
The series (\ref{dn_high_temp_series}) can equivalently be
written as
\beq
D_n = c^{(D)}_{n,n} \, x^n \Big [1 + \sum_{\ell=1}^\infty
  r^{(D)}_{n,\ell} x^\ell \Big ] \ ,
\label{dn_taylor_general}
\eeq
where the ratio $r^{(D)}_{n,j}$ is given by  
\beq
r^{(D)}_{n,\ell} = \frac{c^{(D)}_{n,n+\ell}}{c^{(D)}_{n,n}} \ . 
\label{rdnj}
\eeq

Let us define a variable $t$ as $t=k_>^{-2}$ for $T \ge T_c$ and
$t=k_<^{-2}$ for $T \le T_c$, and auxiliary functions $\sigma_{n,\pm}$ as
follows:
\beq
\sigma_{n,+} = t(t-1)\frac{d\ln D_{n,+}}{dt} - \frac{t}{4}
\label{sigma_plus}
\eeq
and
\beq
\sigma_{n,-} = t(t-1)\frac{d\ln D_{n,-}}{dt} - \frac{1}{4} \ . 
\label{sigma_minus}
\eeq
In \cite{jimbo_miwa}, Jimbo and Miwa showed that the $\sigma_n$
functions are solutions to the following ordinary differential
equation of Painlev\'e VI type (where subscripts $\pm$ on $\sigma_n$
are understood implicitly for $T \ge T_c$ and $T \le T_c$ and
$\sigma_n' \equiv d\sigma_n/dt$):
\beq
    [t(t-1)\sigma_n'']^2 - n^2[(t-1)\sigma_n' - \sigma_n]^2
    + 4\sigma_n'\Big [(t-1)\sigma_n'-\sigma_n- \frac{1}{4} \Big ]
    (t\sigma_n'-\sigma_n)=0 \ . 
\label{p6odeZ}
\eeq
The diagonal correlation functions $D_{n,\pm}$ are then determined in terms
of the $\sigma_{n,\pm}$.  In \cite{ode} with Ghosh, using the result from
\cite{jimbo_miwa}, we derived the general form of the nine terms
beyond the leading term in the high-temperature expansion of $D_n = D_{n,+}$.
(In \cite{ode}, the coefficients $c^{(D)}_{n,j}$ and the ratios
$r^{(D)}_{n,\ell}$ were denoted as $c_{n,j}$ and $r_{n,n+j}$,
respectively.) Our results in \cite{ode} included the following for
the ratios $r^{(D)}_{n,ell}$, in our present notation:
\beq
r^{(D)}_{n,1}=2n
\label{rdn_nplus1}
\eeq
\beq
r^{(D)}_{n,2} = \frac{n(2n^2+3n+5)}{n+1}
\label{rdn_nplus2}
\eeq
\beq
r^{(D)}_{n,3} = \frac{2n(2n^3+5n^2+16n+25)}{3(n+1)}
\label{rdn_nplus3}
\eeq
\beq
r^{(D)}_{n,4} = \frac{4n^6+24n^5+103n^4+372n^3+943n^2+726n-48}{6(n+1)(n+2)}
\label{rdn_nplus4}
\eeq
\beq
r^{(D)}_{n,5} = \frac{4n^7+32n^6+183n^5+930n^4+4031n^3+10228n^2+6972n-960}
{15(n+1)(n+2)} \ ,
\label{rn_nplus5}
\eeq
and so forth for higher-order terms up to $r^{(D)}_{n,9}$. It is interesting
to note that, in an analogous manner, it was possible to use the fact
that correlation functions for the transverse Ising quantum spin chain
at critical field and $T=0$ satisfy a Painlev\'e V equation to derive
a number of properties of the correlation functions for these
functions \cite{mps1}-\cite{msprb},\cite{korepin_de,korepin_book}.


\section{High-Temperature Series for $R_n$}
\label{rn_htseries_appendix}

For reference, in this appendix we list the high-temperature series
that we calculate from our exact results for $R_n$ with $n$ up to 6.
Note that $R_1$ was calculated in \cite{ko}. These series have the
general form of Eq. (\ref{rn_hightemp_series}). We first record some
relations between the elliptic moduli $k_> = 1/k_<$ and the respective
high- and low-temperature expansion variables $v$ and $z$.  The latter
two variables are dual to each other and satisfy
\beq
v=\frac{1-z}{1+z} \ , \quad {\rm equivalently}, \quad z = \frac{1-v}{1+v} \ .
\label{vzrel}
\eeq
Then
\beq
k_> = \bigg [ \frac{2v}{1-v^2} \bigg ]^2
\label{kgval}
\eeq
and
\beq
k_< = \bigg [ \frac{2z}{1-z^2} \bigg ]^2 = \frac{4u}{(1-u)^2} \ .
\label{klval}
\eeq
It it convenient to introduce the rescaled quantities
\beq
\hat k_> = \frac{k_>}{4} \ , \quad \quad \hat k_< = \frac{k_<}{4} \ .
\label{khat}
\eeq

The high-temperature series expansions are 
\beq
R_1 = v + 2 v^3 + 4v^5 + 12v^7 + 42v^9 +164v^{11} +686v^{13}
 +3012v^{15} +O(v^{17})
\label{r1v_taylor}
\eeq
\beq
R_2 = v^2 + 6v^4 + 16v^6 + 46v^8 + 158v^{10} + 618v^{12} + 2618v^{14}
+ 11654v^{16} + O(v^{18})
\label{r2v_taylor}
\eeq
\beq
R_3 = v^3 + 12v^5 + 48v^7 +152v^9 +506v^{11} +1900v^{13} +7902v^{15}
+35114v^{17} + O(v^{19})
\label{r3v_taylor}
\eeq
\beq
R_4 = v^4 + 20v^6 + 118v^8 +452v^{10} +1564v^{12} +5684v^{14}
+22726v^{16} +98708v^{18} + O(v^{20})
\label{r4v_taylor}
\eeq
\beq
R_5 = v^5 + 30v^7 + 250v^9 + 1200v^{11} + 4606v^{13} + 16920v^{15} +
65452v^{17} + 274422v^{19} + O(v^{21})
\label{r5v_taylor}
\eeq
\beq
R_6 = v^6 + 42v^8 + 474v^{10} + 2862v^{12} + 12662v^{14} + 49282v^{16}
+ 189702v^{18} + 770190v^{20} + O(v^{22}) \ .
\label{r6v_taylor}
\eeq
These series can be extended to higher $n$, but these are sufficient to
illustrate our method. 

These high-temperature series for $R_n$ can equivalently be expressed as series
expansions in powers of the variable $\sqrt{k_>}$, using the relation
(\ref{kgval}). However, in contrast to the series in $v$, the series in powers
of $\sqrt{k_>}$ have coefficients that vary in sign, and do not increase
monotonically in magnitude; indeed, some terms have zero coefficients.  
We list these equivalent expansions here. It is convenient to use the 
rescaled quantity $\hat k_>=(1/4)k_>$ as defined Eq. (\ref{khat}), since this 
avoids fractional coefficients.  We have
\beq
R_1 = \hat k_>^{1/2} \bigg [ 1 + \hat k_> + 5\hat k_>^3 - 4\hat k_>^4 
+ 44\hat k_>^5 -60\hat k_>^6 + 469\hat k_>^7 - 820\hat k_>^8 + 
O(\hat k_>^9) \bigg ] 
\label{r1kkg_taylor}
\eeq
\beq
R_2 = \hat k_> \bigg [1 + 4\hat k_> -3\hat k_>^2 + 20\hat k_>^3 
- 24\hat k_>^4 +160\hat k_>^5 -235\hat k_>^6 + 1556\hat k_>^7 
-2568\hat k_>^8 + O(\hat k_>^9) \bigg ]
\label{r2kkg_taylor}
\eeq
\beq
R_3 = \hat k_>^{3/2}\bigg [ 1 + 9\hat k_> -3\hat k_>^2 + 28\hat k_>^3 
+8\hat k_>^4 + 153\hat k_>^5 +233\hat k_>^6 +1008\hat k_>^7 +3588\hat k_>^8 + 
O(\hat k_>^9) \bigg ] 
\label{r3kkg_taylor}
\eeq
\beq
R_4 = \hat k_>^2 \bigg [1 + 16\hat k_> +12\hat k_>^2 +201\hat k_>^4 
- 240\hat k_>^5 +2332\hat k_>^6 -3584\hat k_>^7 + 27280\hat k_>^8 + 
O(\hat k_>^9) \bigg ]
\label{r4kkg_taylor}
\eeq
\beqs
R_5 & = & \hat k_>^{5/2}\bigg [ 1 + 25\hat k_> +60\hat k_>^2 -75\hat k_>^3 
+561\hat k_>^4 -699\hat k_>^5 +4876\hat k_>^6 -5420\hat k_>^7 \cr\cr
&+&45516\hat k_>^8 + O(\hat k_>^9) \bigg ]
\label{r5kkg_taylor}
\eeqs
\beqs
R_6 &=& \hat k_>^3\bigg [ 1 + 36\hat k_> +165\hat k_>^2 -140\hat k_>^3 
+821\hat k_>^4 +276\hat k_>^5 +3092\hat k_>^6 +15440\hat k_>^7 \cr\cr
&-&2484\hat k_>^8 + O(\hat k_>^9) \bigg ] 
\label{r6kkg_taylor}
\eeqs
Note that in the square bracket for $R_1$ in Eq. (\ref{r1kkg_taylor}) there is 
no $\hat k_>^2$ term and in the square bracket for $R_4$ in Eq.
(\ref{r4kkg_taylor}) there is no $\hat k_>^3$ term. 

  For reference, we list numerical values of the $R_n$
for $T \ge T_c$ in Table \ref{rn_high_values}.  For comparison with the
numerical values of $R_n$ as $T \to T_c$, the analytic values of
$(R_n)_{cr}$ with $n$ up to 6 from \cite{row} are as follows:

\beq
(R_1)_{cr} = 2^{-1/2} = 0.707107
\label{r1crit}
\eeq

\beq
(R_2)_{cr}= 1 - \frac{2^2}{\pi^2} =
\bigg (1-\frac{2}{\pi} \bigg ) \bigg (1+\frac{2}{\pi} \bigg ) = 0.594715
\label{r2crit}
\eeq

\beq
(R_3)_{cr}= 2^{3/2} \bigg ( 1 - \frac{2^3}{\pi^2} \bigg ) = 0.53579045
\label{r3crit}
\eeq

\beq
(R_4)_{cr}= 2^4 \bigg ( 1 - \frac{2^4 \cdot 7}{3^2 \pi^2} +
\frac{2^8}{3^2 \pi^4} \bigg ) = 0.497989
\label{r4crit}
\eeq

\beq
(R_5)_{cr}= 2^{15/2} \bigg ( 1 - \frac{2^3 \cdot 19}{3^2 \pi^2} +
\frac{2^9 \cdot 11}{3^4 \pi^4} \bigg ) = 0.470724
\label{r5crit}
\eeq

\beqs
(R_6)_{cr} &=&
2^{12} \bigg ( 1 - \frac{2^2 \cdot 13 \cdot 31}{3 \cdot 5^2 \pi^2}
+ \frac{2^{10} \cdot 7 \cdot 13}{3^3 \cdot 5^2 \cdot \pi^4} -
\frac{2^{22}}{3^6 \cdot 5^2 \pi^6} \bigg ) \cr\cr
&=& 0.449637 \ .
\label{r6crit}
\eeqs
Factorizations of these $(R_n)_{cr}$ for even $n$ were given in
\cite{row}; we have only shown the first of these factorizations, for
$R_2$, here. See also \cite{ayp84}.


\section{Low-Temperature Series for $(R_n)_{\rm conn.}$}
\label{rn_ltseries_appendix}

For reference, we list here the low-temperature series expansions of the
connected correlation functions
$(R_n)_{\rm conn.}$ for $n$ up to 6 here. These have the general form
(\ref{rnconn_smallu}) and are as follows:
\beq
(R_1)_{\rm conn.} = 4u^3 + 28u^4 + 152u^5 + 780u^6 + 3972u^7 + 20348u^8
+ 105192u^9 + 548792u^{10} + O(u^{11})
\label{r1u_conn_taylor}
\eeq
\beq
(R_2)_{\rm conn.} = 4u^4 + 48u^5 + 368u^6 + 2320u^7 + 13428u^8 + 74848u^9
+ 410576u^{10} + 2238496u^{11} + O(u^{12})
\label{r2u_conn_taylor}
\eeq
\beq
(R_3)_{\rm conn.} = 4u^5 + 76u^6 + 832u^7 + 6648u^8 +44852u^9 +276456u^{10}
+1623704u^{11} + 9293292u^{12} + O(u^{13})
\label{r3u_conn_taylor}
\eeq
\beqs
(R_4)_{\rm conn.} &=& 4u^6 + 112u^7 + 1712u^8 + 17584u^9 + 141756u^{10}
+988192u^{11} + 6317392u^{12} \cr\cr
&+& 38365984u^{13} + O(u^{14})
\label{r4u_conn_taylor}
\eeqs
\beqs
(R_5)_{\rm conn.} &=& 4u^7 +156u^8 + 3224u^9 + 42412u^{10} +414228u^{11}
+3331068u^{12} +23619120u^{13} \cr\cr
&+&154485248u^{14} + O(u^{15})
\label{r5u_conn_taylor}
\eeqs
\beqs
(R_6)_{\rm conn.} &=& 4u^8 + 208u^9 + 5632u^{10} + 93680u^{11} +
1111492u^{12} + 10437824u^{13} + 83409104u^{14} \cr\cr
&+& 596805184u^{15} + O(u^{16}) \ .
\label{r6u_conn_taylor}
\eeqs
These series can equivalently be expressed in terms of the variable
$k_<$, using the relation (\ref{klval}). As before, it is convenient to 
use the rescaled variable $\hat k_<=(1/4)k_<$ as defined in Eq. (\ref{khat}), 
since this avoids fractional coefficients.  We have
\beq
(R_1)_{\rm conn.} = 4\hat k_<^3 + 4\hat k_<^4 + 36\hat k_<^5 +52\hat k_<^6
+ 384\hat k_<^7 + 668\hat k_<^8 + 4500\hat k_<^9 + 8820\hat k_<^{10} + 
O(\hat k_<^{11}) 
\label{r1kl_conn_taylor}
\eeq
\beq
(R_2)_{\rm conn.} = 4\hat k_<^4 + 16\hat k_<^5  + 64\hat k_<^6
 + 192\hat k_<^7 + 908\hat k_<^8 + 2256\hat k_<^9 + 12704 \hat k_<^{10} + 
+ O(\hat k_<^{12})  
\label{r2kl_conn_taylor}
\eeq
\beqs
(R_3)_{\rm conn.} &=& 4\hat k_<^5 + 36\hat k_<^6 + 180 \hat k_<^7 
+ 440\hat k_<^8 + 2948 \hat k_<^9 + 5604 \hat k_<^{10} 
+ 42808 \hat k_<^{11} + 74980 \hat k_<^{12} \cr\cr
&+& O(\hat k_<^{13}) \cr\cr
&&
\label{r3kl_conn_taylor}
\eeqs
\beqs
(R_4)_{\rm conn.} &=& 4\hat k_<^6 + 64 \hat k_<^7 + 504 \hat k_<^8 
+ 1344 \hat k_<^9 + 7720 \hat k_<^{10} + 22912 \hat k_<^{11} 
+ 108608 \hat k_<^{12} + 352256 \hat k_<^{13} \cr\cr
&+& O(\hat k_<^{14}) 
\label{r4kl_conn_taylor}
\eeqs
\beqs
(R_5)_{\rm conn.} &=& 4\hat k_<^7 + 100 \hat k_<^8 + 1204 \hat k_<^9 
+ 4900 \hat k_<^{10} + 19224 \hat k_<^{11} + 84708 \hat k_<^{12} + 
311588 \hat k_<^{13} \cr\cr
&+& 1230068 \hat k_<^{14} + O(\hat k_<^{15}) 
\label{r5kl_conn_taylor}
\eeqs
\beqs
(R_6)_{\rm conn.} &=& 4\hat k_<^8 + 144 \hat k_<^9 + 2496 \hat k_<^{10} 
+ 15872 \hat k_<^{11} +58484 \hat k_<^{12} + 250896 \hat k_<^{13} + 
1104448 \hat k_<^{14} \cr\cr
&+& 3668416 \hat k_<^{15} + O(\hat k_<^{16}) \ . 
\label{r6kl_conn_taylor}
\eeqs

Combining Eqs. (\ref{r1u_conn_taylor})-(\ref{r6u_conn_taylor}) with the LT
series expansion for $M^2$, one obtains the LT series expansions for the full
$R_n$ correlation functions:
\beq
R_1 = 1 - 4u^2 -12u^3 - 36u^4 - 120u^5 -448u^6 -1820u^7 -7844u^8
-35256u^9 -163484u^{10} - O(u^{11})
\label{r1u_taylor}
\eeq
\beq
R_2 = 1 - 4u^2 - 16u^3 - 60u^4 - 224u^5 -860u^6 -3472u^7 -14764u^8
-65600u^9 - 301700u^{10} - O(u^{11})
\label{r2u_taylor}
\eeq
\beq
R_3 = 1 - 4u^2 - 16u^3 - 64u^4 - 268u^5 - 1152u^6 - 4960u^7 -21544u^8
-95596u^9 -435820u^{10} - O(u^{11})
\label{r3u_taylor}
\eeq
\beq
R_4 = 1 -4u^2 - 16u^3 - 64u^4 - 272u^5 - 1224u^6 - 5680u^7 - 26480u^8
-122864u^9 - 570520u^{10} - O(u^{11})
\label{r4u_taylor}
\eeq
\beq
R_5 = 1 - 4u^2 - 16u^3 - 64u^4 - 272u^5 - 1228u^6 - 5788u^7 -
28036u^8 - 137224u^9 - 669864u^{10} - O(u^{11})
\label{r5u_taylor}
\eeq
\beq
R_6 = 1 - 4u^2 - 16u^3 - 64u^4 - 272u^5 - 1228u^6 - 5792u^7 - 28188u^8
-140240u^9 - 706644u^{10} - O(u^{11})  \ .
\label{r6u_taylor}
\eeq

The equivalent series expansions, expressed in terms of the variable 
$\hat k_<$, are
\beq
R_1 = 1 - 4\hat k_<^2 + 4\hat k_<^3 - 20\hat k_<^4 + 36\hat k_<^5
-172\hat k_<^6 + 384\hat k_<^7 -1796\hat k_<^8 + 4500\hat k_<^9 
-20748 \hat k_<^{10} + O(\hat k_<^{11}) 
\label{r1kl_taylor}
\eeq
\beq
R_2 = 1 - 4\hat k_<^2 -20\hat k_<^4 +16\hat k_<^5
 -160\hat k_<^6 +192\hat k_<^7 -1556\hat k_<^8 +2256\hat k_<^9 
 -16864\hat k_<^{10} -27392\hat k_<^{10} + O(\hat k_<^{11}) 
\label{r2kl_taylor}
\eeq
\beq
R_3 = 1 - 4\hat k_<^2 - 24 \hat k_<^4 + 4 \hat k_<^5 - 188 \hat k_<^6 
+ 180 \hat k_<^7 - 2024\hat k_<^8 + 2948 \hat k_<^9 - 23964 \hat k_<^{10} 
+ O(\hat k_<^{11}) 
\label{r3kl_taylor}
\eeq
\beq
R_4 = 1 - 4\hat k_<^2 -24 \hat k_<^4 - 220 \hat k_<^6 + 64 \hat k_<^7 
- 1960 \hat k_<^8 + 1344 \hat k_<^9 - 21848 \hat k_<^{10} +O(\hat k_<^{11}) 
\label{r4kl_taylor}
\eeq
\beq
R_5 = 1 - 4\hat k_<^2 - 24 \hat k_<^4 - 224 \hat k_<^6 + 4 \hat k_<^7 
- 2364 \hat k_<^8 + 1204 \hat k_<^9 -24668 \hat k_<^{10} + O(\hat k_<^{11}) 
\label{r5kl_taylor}
\eeq
\beq
R_6 = 1 - 4\hat k_<^2 - 24 \hat k_<^4 - 224 \hat k_<^6 -2460 \hat k_<^8 
+ 144 \hat k_<^9 - 27072 \hat k_<^{10} + O(\hat k_<^{11})  \ . 
\label{r6kl_taylor}
\eeq

We list numerical values of the $R_n$
for $T \le T_c$ in Table \ref{rn_low_values}.


\end{appendix} 



\begin{table}
  \caption{\footnotesize{Numerical values of the $R_n$ for $1 \le n \le 6$
      and $T \ge T_c$ as functions of $k_>$. For reference, the
      values of $v$ and $T/T_c$ corresponding to each value of $k_>$
      are also shown. In this and the other tables, the notation
      $a$e-$n$ means $a \times 10^{-n}$. }}
  \begin{center}
\begin{tabular}{|c|c|c||c|c|c|c|c|c|} \hline\hline
$k_>$ & $v$ & $T/T_c$ & $R_1$ & $R_2$ & $R_3$ & $R_4$ & $R_5$ & $R_6$ \\
\hline
0   & 0     & $\infty$& 0      & 0      & 0        & 0         & 0
& 0         \\

0.1 & 0.1543  & 8.828 & 0.1621 & 0.02746& 0.4837e-2&0.8797e-3 & 1.642e-4
& 0.3127e-4 \\

0.2 & 0.2134  & 4.435 & 0.2349 & 0.05974& 0.01617 & 0.4578e-2 & 1.338e-3
& 0.4000e-3 \\

0.3 & 0.2559  & 2.981 & 0.2950 & 0.09684& 0.03431 & 0.01279 & 0.4926e-2
& 1.939e-3 \\

0.4 & 0.2897  & 2.260 & 0.3494 & 0.1389 & 0.06011 & 0.02740 & 0.01290
& 0.6200e-2 \\

0.5 & 0.3178  & 1.832 & 0.4013 & 0.1864 & 0.09463 & 0.05059 & 0.02790
& 0.01569 \\

0.6 & 0.3420  & 1.549 & 0.4525 & 0.2400 & 0.13935  & 0.08507 & 0.05348
& 0.03426   \\

0.7 & 0.3632  & 1.350 & 0.5045 & 0.3011 & 0.1965  & 0.1345 & 0.09462
& 0.06774   \\

0.8 & 0.3820  & 1.203 & 0.5595 & 0.3723 & 0.2700  & 0.2046 & 0.1590
& 0.1256   \\

0.9 & 0.3989  & 1.090 & 0.6210 & 0.4596 & 0.3684  & 0.3070 & 0.2617
& 0.2262   \\

1   & 0.4142  & 1     & 0.7071 & 0.5947 & 0.5358  & 0.4980 & 0.4707
& 0.4496   \\
\hline\hline
\end{tabular}
\end{center}
\label{rn_high_values}
\end{table}


\begin{table}
  \caption{\footnotesize{Numerical values of the $R_n$
      for $1\le n \le 6$ and $T \le T_c$ as functions of $k_<$ For
      reference, the values of $z$ and $T/T_c$ corresponding to each
      value of $k_<$ are also shown.}}
\begin{center}
\begin{tabular}{|c|c|c||c|c|c|c|c|c|} \hline\hline
$k_<$ & $z$ & $T/T_c$ & $R_1$ & $R_2$ & $R_3$ & $R_4$ & $R_5$ & $R_6$ \\
\hline
0  & 0      & 0      & 0      & 0      & 0      & 0      & 0      & 0      \\

0.1& 0.1543 & 0.2940 & 0.9976 & 0.9975 & 0.9975 & 0.9975 & 0.9975 & 0.9975 \\

0.2& 0.2134 & 0.3811 & 0.9904 & 0.9800 & 0.9899 & 0.9898 & 0.9898 & 0.9898 \\

0.3& 0.2559 & 0.4593 & 0.9786 & 0.9769 & 0.9767 & 09767. & 0.9767 & 0.9767 \\

0.4& 0.2897 & 0.5351 & 0.9622 & 0.9580 & 0.95745& 0.9574 & 0.95735&0.95735 \\

0.5& 0.3178 & 0.6105 & 0.9410  & 0.9325& 0.9310 & 0.9307 & 0.9306 & 0.9306 \\

0.6& 0.3420 & 0.6865 & 0.9144  & 0.8992& 0.8958 & 0.8949 & 0.8946 & 0.8945  \\

0.7& 0.3632 & 0.7634 & 0.8817  & 0.8562 & 0.8491& 0.8467 & 0.8458 & 0.8454 \\

0.8& 0.3820 & 0.8413 & 0.8412 & 0.8003 & 0.7864 & 0.78065& 0.7779 & 0.7765 \\

0.9& 0.3989 & 0.9202 & 0.7893 & 0.7245 & 0.6978 & 0.6842 & 0.6764 & 0.6716 \\

1  & 0.4142 & 1      & 0.7071 & 0.5947 & 0.5358 & 0.4980 & 0.4707 & 0.4496 \\
\hline\hline
\end{tabular}
\end{center}
\label{rn_low_values}
\end{table}


\end{document}